\documentclass[twocolumn]{emulateapj}
\usepackage{graphicx,amsmath,amssymb}% Include figure files
\usepackage{bm,url,color,euscript}% bold math
\usepackage{natbib}
\definecolor{darkblue}{rgb}{.0, .0,.6}
\definecolor{brown}{rgb}{0.5,0.2,0.2}
\definecolor{dgreen}{rgb}{0.15,0.6,0.15}
\definecolor{gray1}{rgb}{0.8,0.8,0.8}
\definecolor{dyellow}{rgb}{0.6,0.6,0.1}
\definecolor{lblue}{rgb}{0.1,0.1,0.8}
\newcommand{\beq}{\begin{equation}}
\newcommand{\eeq}{\end{equation}}

\newcommand{\bfb}{\mbox{\boldmath $b$}}
\newcommand{\bff}{\mbox{\boldmath $f$}}
\newcommand{\bfF}{\mbox{\boldmath $F$}}

\newcommand{\bfk}{\mbox{\boldmath $k$}}

\newcommand{\bfv}{\mbox{\boldmath $v$}}
\newcommand{\bfx}{\mbox{\boldmath $x$}}

\newcommand{\bfB}{\mbox{\boldmath $B$}}
\newcommand{\bfJ}{\mbox{\boldmath $J$}}

\newcommand{\ex}{\mbox{{\boldmath $e$}}_{1}}
\newcommand{\ey}{\mbox{{\boldmath $e$}}_{2}}
\newcommand{\ez}{\mbox{{\boldmath $e$}}_{3}}
\newcommand{\bfemf}{\mbox{\boldmath ${\cal E}$}}
\newcommand{\bnabla}{\mbox{\boldmath $\nabla$}}
\newcommand{\cross}{\mbox{\boldmath $\times$}}
\newcommand{\cendot}{\mbox{\boldmath $\cdot\,$}}
\newcommand{\rem}{{\rm Rm}}
\newcommand{\re}{{\rm Re}}
\newcommand{\pr}{{\rm Pr}}
\newcommand{\sh}{{\rm S_h}}
\newcommand{\ssc}{\scriptsize}
\shorttitle{Numerical studies of dynamo action in a turbulent shear flow -- I}
\shortauthors{Singh \& Jingade}
\begin{document}
\title{Numerical studies of dynamo action in a turbulent shear flow -- I}
\author{Nishant K. Singh\altaffilmark{1,2,3,4} and
Naveen Jingade\altaffilmark{5}}
\email{nishant@nordita.org}
\email{naveenjingade@physics.iisc.ernet.in}
\altaffiltext{1}{Raman Research Institute, Sadashivanagar,
Bangalore 560 080, India}
\altaffiltext{2}{Joint Astronomy Programme, Indian Institute
of Science, Bangalore 560 012, India}
\altaffiltext{3}{Inter-University Centre for Astronomy and
Astrophysics, Post Bag 4, Ganeshkhind, Pune 411 007, India}
\altaffiltext{4}{Nordita, KTH Royal Institute of Technology
and Stockholm University,
Roslagstullsbacken 23, SE-10691 Stockholm, Sweden}
\altaffiltext{5}{Indian Institute of Science, Bangalore 560 012, India}

\date{\today}

\begin{abstract}
We perform numerical experiments to study the shear dynamo problem where
we look for the growth of large--scale magnetic field due to non--helical
stirring at small scales in a background linear shear flow, in previously
unexplored parameter regimes. We demonstrate the large--scale
dynamo action in the limit when
the fluid Reynolds number ($\re$) is below unity whereas the magnetic
Reynolds number ($\rem$) is above unity; the exponential growth rate scales
linearly with shear, which is consistent with earlier numerical works.
The limit of low $\re$ is particularly
interesting, as seeing the dynamo action in this limit would provide enough
motivation for further theoretical investigations, which may focus
the attention to this analytically more tractable limit of
$\re < 1$ as compared to more formidable limit of $\re > 1$.
We also perform simulations in the regimes
when, (i) both ($\re$, $\rem$) $< 1$; (ii) $\re > 1$ \& $\rem < 1$,
and compute all components of the turbulent transport coefficients
($\alpha_{ij}$ and $\eta_{ij}$) using the test--field method.
A reasonably good agreement is seen between our results
and the results of earlier analytical works \citep{SS10,SS11} in
the similar parameter regimes.
\end{abstract}


\maketitle

\section{Introduction}

Magnetic fields observed in various astrophysical systems such as, the
Earth, the Sun, the disc galaxies, accretion discs etc., possess
large--scale magnetic fields in addition to a fluctuating component.
The magnetic field survives for time scales much larger than the
diffusion time scales in those systems,
and therefore are thought to be self--sustained by turbulent
dynamo action. The standard model of such a turbulent dynamo
to produce large--scale magnetic field involves amplification
of seed magnetic fields due to the usual $\alpha$--effect,
where $\alpha$ is a measure of net kinetic helicity in the flow
(see e.g. \citet{Mof78,Par79,KR80,BS05,BSS12}).
As it is not necessary that the turbulent flow be always helical,
it is interesting to study the dynamo action in non--helically
forced shear flows. Dynamo action due to shear and turbulence, in
the absence of the $\alpha$--effect,
received some attention in the astrophysical contexts of accretion
discs \citep{VB97} and galactic discs \citep{Bla98,SurSub09}.
The presence of large--scale shear in turbulent flows is expected
to have significant effects on transport properties
\citep{RK06,RS06,LK09,SS10,SS11}.
It has also been demonstrated that the mean shear in conjunction
with the rotating turbulent convection gives rise to the growth
of large--scale magnetic fields \citep{KKB08,HP09}.
The problem of our interest may be stated as follows: in the absence of
the $\alpha$--effect, will it be possible to generate large--scale magnetic
field just due to the action of non--helical turbulence in
background shear flow on the seed magnetic field?
This question just posed was studied numerically in the recent past by
\cite{BRRK08,You08a,You08b}. These works
clearly demonstrated the growth of large--scale magnetic fields
due to non--helical stirring at small scale in the background linear
shear flow.

Although various mechanisms have been proposed to resolve the shear
dynamo problem, it is still not clear what really drives the dynamo
action in such systems. The presence of the magnetic helicity
flux could be a candidate for the growth of
large--scale magnetic field \citep{VC01,BS05,SV11}.
Yet another possibility that has been suggested is the shear--current effect
\citep{RK03,RK04,RK08}, where the shear--current term in the expression for
the mean electromotive force (EMF) is thought to generate the cross--shear
component of mean magnetic field from the shearwise component.
However, some analytic calculations \citep{RS06,RK06,SS09a,SS09b,SS10,SS11}
and numerical experiments \citep{BRRK08} find that the sign of the
shear--current term is unfavorable for dynamo action.
Quasilinear kinematic theories of \cite{SS09a,SS09b}, and low magnetic
Reynolds number ($\rem$) theories of \cite{SS10,SS11} found no evidence of
dynamo action; in these works, a Galilean--invariant formulation of the shear dynamo problem
was developed, in which the $\alpha$ effect was strictly zero, and unlike earlier works, the
shear was treated non-perturbatively. 
It has been discussed that the mean magnetic field could grow due to a process
known as the incoherent alpha--shear mechanism, in which, the fluctuations in
$\alpha$ with no net value, together with the mean shear might drive the large--scale
dynamo action \citep{VB97,Sok97,Sil00,Pro07,KR08,BRRK08,SurSub09,RP12,SS14}.
Recent analytical works by \cite{HMS11,M11,MB12} predict the growth
of mean--squared magnetic field by considering fluctuating $\alpha$ in
background shear in the limit of small Reynolds numbers.
\cite{SS14} discuss the possibility of the growth of mean magnetic field in shearing background
by considering zero--mean temporal fluctuations in $\alpha$, which have finite correlation times.

It should be noted that all the earlier numerical experiments done so far have been carried
out for both the fluid Reynolds number ($\re$) and the magnetic Reynolds number ($\rem$) above
unity, the limit for which rigorous theory explaining 
the origin of \emph{the shear dynamo} is yet to come.
In order to make step-by-step progress analytically, it
seems necessary to explore the regime, $\re < 1$ and $\rem > 1$ before one aims to have
a theory which is valid for both ($\re$, $\rem$) $> 1$. Such thoughts motivated us to look
for numerical experiment carried out in the regime when $\re < 1$ and $\rem > 1$. 

In this paper, we present numerical simulations for the shear dynamo problem which can be broadly
classified in following three categories: (i) The regime when both $\re$ and $\rem$ are
less than unity. This is done for comparison with earlier analytical work \citep{SS11};
(ii) $\re > 1$ and $\rem < 1$; and (iii) the regime when $\re < 1$ and $\rem > 1$. We have used the
Pencil Code \footnote{See http://www.nordita.org/software/pencil-code.} for all the simulations
presented in this paper and followed the method given in \cite{BRRK08}.
In \textsection~2 we begin with the fundamental equations of
magnetohydrodynamics in a background linear shear flow. We then
consider the case when the mean--magnetic field is a function only of the spatial coordinate $x_3$ and time $t$. We
briefly describe the transport coefficients and discuss the test field method. Few important details of the
simulation are presented. In \textsection~3, we put together all the results in
three parts, namely, part A, part B and part C corresponding to the three categories discussed above. We also
make comparisons with analytical works of \cite{SS10,SS11}.
In \textsection~4, we present our conclusions.

\section{The model and numerical set up}

Let $(\ex,\ey,\ez)$ be the unit basis vectors of a Cartesian coordinate system in the laboratory 
frame. Using notation $\bfx = (x_1,x_2,x_3)$ for the position vector and $t$ for time, we write 
the total fluid velocity as $(Sx_1\ey + \bfv)$, where $S$ is the rate of shear parameter 
and $\bfv(\bfx, t)$ is the velocity deviation from the background shear flow. Let $\bfB^{\rm tot}$
be the total magnetic field which obeys the induction equation. We have performed 
numerical simulations using the Pencil Code which is a publicly available code suited for
weakly compressible hydrodynamic flows with magnetic fields. We consider
velocity field $\bfv$ to be compressible and write the momentum, continuity and induction equations for a 
compressible fluid of mass density $\rho$:

\begin{eqnarray}
\label{NS}
&&\left( \frac{\partial}{\partial t} + S x_1 \frac{\partial}{\partial x_2}
\right) \bfv + Sv_1\ey + (\bfv \cendot \bnabla)\bfv =
-\frac{1}{\rho}\bnabla P \nonumber \\[1ex]
&&\hskip1.2in +\; \frac{\bfJ^{\rm tot} \cross 
\bfB^{\rm tot}}{\rho} + \bfF_{\rm visc} + \bff \\[2ex]
\label{cont}
&&\left( \frac{\partial}{\partial t} + S x_1 \frac{\partial}{\partial x_2}
\right)\rho + (\bfv \cendot \bnabla)\rho = -\rho \bnabla \cendot \bfv \\[2ex]
\label{ind}
&&\left( \frac{\partial}{\partial t} + S x_1 \frac{\partial}{\partial x_2}
\right)\bfB^{\rm tot}-SB_1^{\rm tot}\ey =
\bnabla\cross(\bfv \cross \bfB^{\rm tot}) \nonumber \\[1ex]
&&\hskip2.2in
+\; \eta \nabla^2 \bfB^{\rm tot}
\end{eqnarray}

\noindent
where $\bfF_{\rm visc}$ denotes the viscous term, $\bff$ is the random stirring force 
per unit mass and we write $\bfJ^{\rm tot} = (\bnabla \cross \bfB^{\rm tot})$, for simplicity,
instead of the usual definition $\bfJ^{\rm tot} = (\bnabla \cross \bfB^{\rm tot})/\mu_0$.
$\mu_0$ and $\eta$ represent the magnetic permeability and magnetic diffusivity, respectively.
Our aim is to investigate the case of \emph{incompressible} magnetohydrodynamics in a background linear 
shear flow with a non--helical random forcing at small scales. In order to do that with Pencil Code, we
limit ourselves to the cases for which the root--mean--squared velocity, $v_{\rm rms}$, is small
compared with the sound speed, making the Mach number (${\rm Ma}$)
very small. In this case the solutions of
compressible equations approximate the solutions of incompressible equations. When the velocity
field $\bfv$ is incompressible (or weakly compressible), the viscous term in Eqn.~(\ref{NS}) becomes
$\bfF_{\rm visc}=\nu \nabla^2 \bfv$ ($\nu$ denotes the coefficient of kinematic viscosity) and the
right hand side of continuity equation vanishes.

\subsection{Mean--field induction equation}

Various transport phenomena have traditionally been studied in the framework of
mean--field theory \citep{Mof78,KR80,BS05}. Applying Reynolds averaging to the 
induction Eqn.~(\ref{ind}) we find that the mean magnetic field, 
$\bfB(\bfx, t)$, obeys the following (mean--field induction) equation:

\beq
\left(\frac{\partial}{\partial t} \;+\; Sx_1\frac{\partial}{\partial x_2}\right)\bfB \;-\; SB_1\ey \;=\; 
\bnabla\cross\bfemf \;+\; \eta\bnabla^2\bfB
\label{meanindeqn}
\eeq

\noindent
where $\eta$ is the microscopic resistivity, and  $\bfemf$ is the mean electromotive 
force (EMF), $\bfemf = \left<\bfv'\cross\bfb'\right>$, where $\bfv'$ and $\bfb'$ are the 
fluctuations in the velocity and magnetic fields, respectively. 
We perform numerical simulations in a cubic domain of size $L\times L\times L$, where the
mean--field $Q$ of some quantity $Q^{\rm tot}$ is defined by

\beq
Q(x_3, t) \;=\; \frac{1}{L^2} \int_{-L/2}^{L/2}\,\int_{-L/2}^{L/2}\,
Q^{\rm tot}(x_1, x_2, x_3, t)\, \mathrm{d}x_1\, \mathrm{d}x_2
\label{mean}
\eeq

\noindent
Thus the mean--field quantities discussed here are functions of $x_3$ and time $t$.
The mean EMF is, in general, a 
\emph{functional} of the mean magnetic field, $B_l$. For a slowly varying 
mean magnetic field, the mean EMF can approximately be written as a \emph{function} 
of $B_l$ and $B_{lm}$; see \citep{BRRK08,SS11}:

\beq
{\cal E}_i \;=\; \alpha_{il}(t) B_l(\bfx,t) \;-\; \eta_{iml}(t)\,\frac{\partial B_l(\bfx,t)}{\partial x_m}
\label{emfslow}
\eeq

\noindent
where $\alpha_{il}(t)$ and $\eta_{iml}(t)$ are the \emph{transport coefficients}, which evolve in time in the beginning
and saturate at late times.

\subsection{Transport coefficients}

Previous studies have shown that the mean value of $\alpha_{il}$ is
zero so long as the stirring is 
non--helical \citep{BRRK08,SS09a,SS09b,SS10,SS11}, but it shows
zero--mean temporal fluctuations in simulations.
With the definition of the mean--field as given in Eqn.~(\ref{mean}),
we note that the mean magnetic field $\bfB =\bfB(x_3, t)$. The condition 
$\bnabla \cendot\bfB = 0$ implies that $B_3$ is uniform in space, and it can be 
set to zero; hence we have $\bfB = (B_1, \,B_2, \,0)$. Thus, Eqn.~(\ref{emfslow}) 
for the mean EMF gives $\bfemf = ({\cal E}_1, \,{\cal E}_2,\, 0)$, with 

\beq
{\cal E}_i = \alpha_{ij} B_j-\,\eta_{ij}\,J_j\,;\;\;
\bfJ \;=\; \bnabla\cross \bfB = \left( -\frac{\partial B_2}{\partial x_3},
\;\frac{\partial B_1}{\partial x_3},\;0 \right) 
\label{emftf}
\eeq

\noindent
where all components of $\alpha_{ij}$ show zero--mean temporal fluctuations as the forcing is non--helical
and $2$--indexed magnetic diffusivity tensor $\eta_{ij}$ has four components, 
$(\eta_{11}, \,\eta_{12}, \,\eta_{21}, \,\eta_{22})$, which are defined in terms 
of the $3$--indexed object $\eta_{iml}$ by

\beq
\eta_{ij} = \epsilon_{lj3}\,\eta_{i3l}\,;\,\mbox{which implies,}
\;\eta_{i1} = -\,\eta_{i32}\,;\,\eta_{i2} = \eta_{i31}.
\label{etatf}
\eeq

\noindent
Substituting Eqn.~(\ref{emftf}) for $\bfemf$ in Eqn.~(\ref{meanindeqn}), we get the
evolution equation for the mean magnetic field.
The diagonal components, $\eta_{11}$ and $\eta_{22}$, augment the 
microscopic resistivity, $\eta$, whereas the off--diagonal components, 
$\eta_{12}$ and $\eta_{21}$, lead to cross--coupling of  $B_1$ and $B_2$.
It was shown in \cite{SS11} that each component of $\eta_{ij}$ starts from zero
at time $t=0$ and saturates at some constant value ($\eta^{\infty}_{ij}$)
at late times. Here we aim to measure these saturated quantities.

\subsection{Test field method}

We use test field method to determine the transport coefficients $\alpha_{ij}$ and $\eta_{ij}$.
The procedure has been described in detail in \cite{BRRK08} (see also references
therein). A brief description of the method is as follows: Let $\bfB^q$ be a set of
test--fields and $\bfemf^q$ be the EMF corresponding to the test field $\bfB^q$.
Subtracting Eqn.~(\ref{meanindeqn}) from Eqn.~(\ref{ind}), we get the
evolution equation for the fluctuating field $\bfb'$. With properly chosen $\bfB^q$
and the flow $\bfv'$, we can numerically solve for the fluctuating field $\bfb^q$.
This enables us to determine $\bfemf^q$ which can then be used to find $\alpha_{ij}$ and $\eta_{ij}$ 
using ${\cal E}_i^q = \alpha_{ij}\,B_j^q-\,\eta_{ij}\,J_j^q$ where $\bfJ^q = \bnabla\cross \bfB^q$.

There could be various choices for the number and form of the test fields which 
essentially depends on the problem that one is trying to solve. For our purposes,
we have chosen the test fields, denoted as $\bfB^{qc}$, defined by,

\beq
\bfB^{1c} = B\left(\cos[k x_3], 0, 0\right)\,;\quad
\bfB^{2c} = B\left(0, \cos[k x_3], 0\right)
\label{tfBcos} 
\eeq

\noindent where $B$ and $k$ are assumed to be constant. Using Eqn.~(\ref{tfBcos})
in the expression ${\cal E}_i^q = \alpha_{ij}\,B_j^q\,-\,\eta_{ij}\,J_j^q$, we find the corresponding 
mean EMF denoted by $\bfemf^{qc}$ as,

\begin{eqnarray}
{\cal E}_i^{1c} &=& \alpha_{i1}\,B\,\cos[k x_3] + 
\eta_{i2}\, B k \,\sin[k x_3] \nonumber \\[2ex]
{\cal E}_i^{2c} &=& \alpha_{i2}\,B\,\cos[k x_3] -
\eta_{i1}\, B k \,\sin[k x_3]\,;\; i = 1,2
\label{tfemfcos}
\end{eqnarray}

\noindent
Here we have four equations but eight unknowns $(\eta_{11},... \eta_{22}\,;\,\alpha_{11},...\alpha_{22})$. So, we
further consider the following set of test field denoted as $\bfB^{qs}$ defined by,

\beq
\bfB^{1s} = B\left(\sin[k x_3], 0, 0\right)\,;\;\;
\bfB^{2s} = B\left(0, \sin[k x_3], 0\right)
\label{tfBsin} 
\eeq

\noindent where $B$ and $k$ are assumed to be constant as before. 
Using Eqn.~(\ref{tfBsin}) in the expression ${\cal E}_i^q = \alpha_{ij}\,B_j^q\,-\,\eta_{ij}\,J_j^q$, 
we find the corresponding mean EMF denoted by $\bfemf^{qs}$ as,

\begin{eqnarray}
{\cal E}_i^{1s} &=& \alpha_{i1}\,B\,\sin[k x_3] -
\eta_{i2}\, B k \,\cos[k x_3] \nonumber \\[2ex]
{\cal E}_i^{2s} &=& \alpha_{i2}\,B\,\sin[k x_3] +
\eta_{i1}\, B k \,\cos[k x_3]\,;\; i = 1,2
\label{tfemfsin}
\end{eqnarray}

\noindent Using Eqns.~(\ref{tfemfcos}) and (\ref{tfemfsin}) we can write,

\begin{eqnarray}
\label{tfalpij}
\alpha_{i1} &=& \frac{1}{B} \left( {\cal E}_i^{1c} \cos[k x_3] +
{\cal E}_i^{1s} \sin[k x_3] \right) \nonumber \\[2ex]
\alpha_{i2} &=& \frac{1}{B} \left( {\cal E}_i^{2c} \cos[k x_3] +
{\cal E}_i^{2s} \sin[k x_3] \right)\;;\; i=1,2 \\[2ex]
\eta_{i1} &=& -\frac{1}{B k} \left( {\cal E}_i^{2c} \sin[k x_3] -
{\cal E}_i^{2s} \cos[k x_3] \right) \nonumber \\[2ex]
\eta_{i2} &=& \frac{1}{B k} \left( {\cal E}_i^{1c} \sin[k x_3] -
{\cal E}_i^{1s} \cos[k x_3] \right)\;;\; i=1,2
\label{tfetaij}
\end{eqnarray}

\noindent Thus from the Eqns.~(\ref{tfalpij}) and (\ref{tfetaij}) we can determine the unknown 
quantities $\alpha_{ij}$ and $\eta_{ij}$. For homogeneous turbulence being considered here,
transport coefficients need to be independent of $x_3$, therefore, the apparent dependence
on $x_3$ through the terms $\sin[k x_3]$ and $\cos[k x_3]$ in Eqns.~(\ref{tfalpij}) and (\ref{tfetaij})
have to be compensated by $x_3-$dependent ${\cal E}_i$'s given
by Eqns.~(\ref{tfemfcos}) and (\ref{tfemfsin}).

We use ``shear--periodic'' boundary conditions to solve Eqns.~(\ref{NS}--\ref{ind}) in the same manner 
as given in \cite{BRRK08}. Shear--periodic boundary conditions have been widely used in 
numerical simulations of a variety of contexts. Simulations of local patches of planetary rings \citep{WT88},
local dynamics of differentially rotating discs in astrophysical systems \citep{BH98, BT08},
nonlinear evolution of perturbed shear flow in two--dimensions with the ultimate goal to understand 
the dynamics of accretion disks \citep{Lit07}, the shear dynamo 
\citep{BRRK08,You08a,You08b,KKB08} etc are few examples.

The random forcing function $\bff$ in Eqn.~(\ref{NS}) is assumed to be \emph{non--helical}, 
\emph{homogeneous}, \emph{isotropic} and \emph{delta--correlated--in--time}. Further, we assume that
the vector function $\bff$ is solenoidal and the forcing is confined to a spherical shell of magnitude
$|\bfk_f| = k_f$ where the wavevector $\bfk_f$ signifies the energy--injection scale ($l_f = 2\pi /k_f$) 
of turbulence. This can be \emph{approximately} achieved by following the method described in \cite{BRRK08}.
We note that although the random forcing $\bff$ is delta--correlated--in--time, the resulting fluctuating 
velocity field $\bfv$ will \emph{not} be delta--correlated--in--time (this is due to the inertia as has 
been pointed out in \cite{BRRK08}). This has been rigorously proved
in \cite{SS11} in the limit of small fluid Reynolds number, the limit which we aim to explore in the
present manuscript. Another important fact to note is that in the limit of small $\re$ the non--helical 
forcing has been shown to give rise to non--helical velocity field in the reference \cite{SS11}; whether
this is true even in the limit of high $\re$ has not been proved yet. Thus performing the simulation
in the limit $\re<1$ with non--helical forcing guarantees the fact that \emph{the fluctuating
velocity field is also non--helical}. 

\section{Results and Discussion}

We have explored following three parameter regimes:
(i) $\re < 1$ and $\rem < 1$;  (ii) $\re > 1$ and $\rem < 1$;
(iii) $\re < 1$ and $\rem > 1$. All the results obtained
in numerical simulations for various parameter regimes are being presented.
As all the transport coefficients show temporal fluctuations
about some constant value, we take long time averages of the quantities
and denote them by $\eta_{ij}^\infty$. The turbulent diffusivity, $\eta_t$,
is defined in terms of components of magnetic diffusivity tensor as follows:
\beq
\eta_t \;=\; \frac{1}{2}(\eta^{\infty}_{11} \,+\, \eta^{\infty}_{22})\;,\qquad \eta_T \;=\; \eta \,+\, \eta_t\;,\qquad
\label{etas}
\eeq
We note that the rate of shear parameter, $S<0$, and $K$ is the smallest finite
wavenumber in the $x_3$-direction.
We now define various dimensionless quantities: The \emph{fluid Reynolds 
number}, $\re = v_{\rm rms}/(\nu k_f)\,$; the \emph{magnetic Reynolds number}, 
$\rem = v_{\rm rms}/(\eta k_f)\,$; the \emph{Prandtl number}, ${\pr} = \nu/\eta\,$;
the dimensionless \emph{Shear parameter}, ${\sh} = S/(v_{\rm rms}k_f)\,$.
Symbols used in these definitions have usual meanings.

\begin{figure}%[t]
\begin{center}
\includegraphics[scale=0.43,angle=-90]{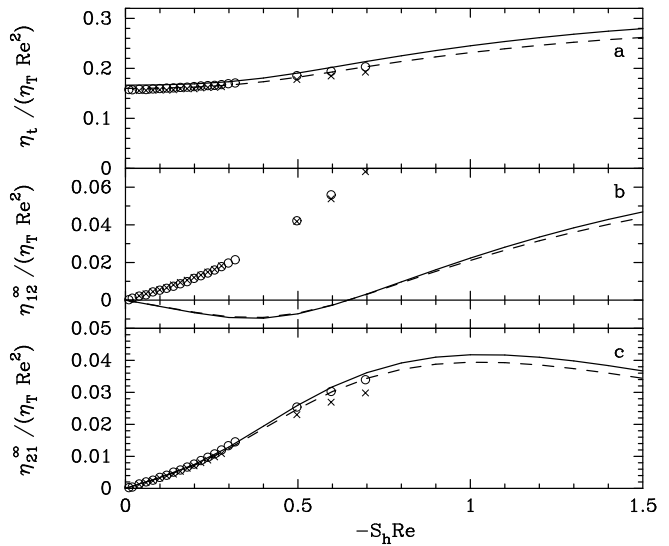}
\caption{Plots of the saturated quantities $\eta_t$,
$\eta^{\infty}_{12}$ and $\eta^{\infty}_{21}$
for $\re=\rem \approx 0.16$, and $\re=\rem \approx 0.46$
(i.e. $\pr=1$), versus the dimensionless parameter $(-\sh \re)$.
Comparison of the results of the simulations with the theory presented
in \cite{SS11} is shown, where the lines (`bold' and `dashed')
correspond to the theory, whereas the symbols
(`$\circ$' and `$\times$') correspond to the simulations.
The `bold' lines and the symbols `$\circ$' are for
$\re=\rem \approx 0.16$, whereas the `dashed' lines
and the symbol `$\times$' are for $\re=\rem \approx 0.46$.}
\label{tfnij_th_Pr1}
\end{center}
\end{figure}

\begin{center}
 {\bf PART A: $\re < 1$ and $\rem < 1$}
\end{center}
It is a necessary step to compare the numerical results obtained in
this parameter regime with the earlier analytical work in which the
general functional form for the saturated values of magnetic
diffusivities, $\eta_{ij}$, was predicted (see Eqn.~(60) and related
discussion in \cite{SS11}). It is useful to recall the following
expression for the growth rate of the mean magnetic field, obtained
from the mean field theory (see e.g. \citet{BRRK08,SS11}):
\beq
\frac{\lambda_{\pm}}{\eta_T \,K^2} \;=\; -1\; \pm \, \frac{1}{\eta_T} \sqrt{\eta^{\infty}_{21}
\left( \frac{S}{K^2} + \eta^{\infty}_{12} \right) \,+\,\epsilon^2}
\label{roots}
\eeq
where,
\beq
\epsilon \;=\; \frac{1}{2}(\eta^{\infty}_{11} \,-\, \eta^{\infty}_{22})\;;
\quad \mbox{and} \quad S\,<\,0
\eeq

\begin{figure}%[t]
\begin{center}
\includegraphics[scale=0.43,angle=-90]{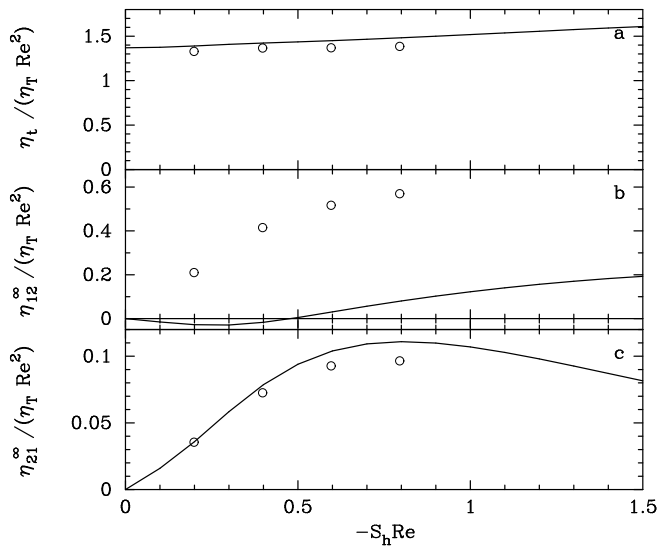}
\caption{Same as Fig.~(\ref{tfnij_th_Pr1}), but for $\re \approx 0.13$
and $\rem \approx 0.64$ (i.e. $\pr=5$), versus the dimensionless
parameter $(-\sh \re)$. The bold lines correspond to the theory, whereas
the symbols `$\circ$' correspond to the simulations.}
\label{tfnij_th_Pr5}
\end{center}
\end{figure}

\begin{figure}%[t]
\begin{center}
\includegraphics[scale=0.43,angle=-90]{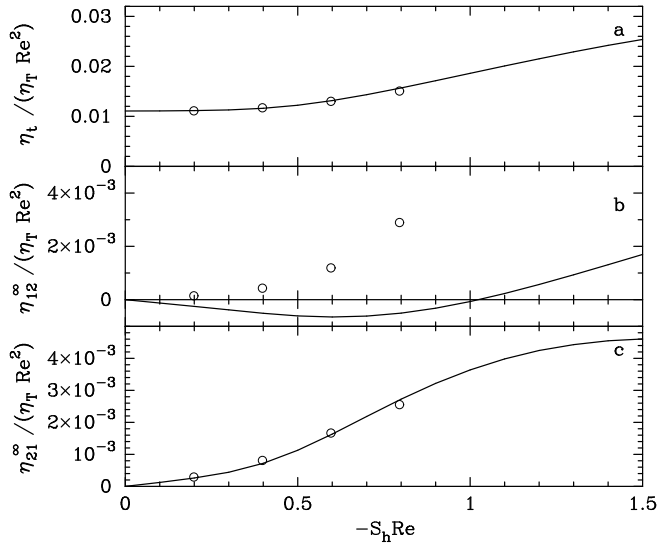}
\caption{Same as Fig.~(\ref{tfnij_th_Pr1}), but for $\re \approx 0.13$
and $\rem \approx 0.025$ (i.e. $\pr=0.2$), versus the dimensionless
parameter $(-\sh \re)$. The bold lines correspond to the theory, whereas
the symbols `$\circ$' correspond to the simulations.}
\label{tfnij_th_Prpt2}
\end{center}
\end{figure}

\noindent Figures~(\ref{tfnij_th_Pr1}--\ref{tfnij_th_Prpt2}) display
plots of $\eta_t$, $\eta^{\infty}_{12}$ and $\eta^{\infty}_{21}\,$, 
versus the dimensionless parameter $\left(-\sh \re\right)$, which demonstrate
the comparison of the results from a direct numerical simulation with $64^3$ mesh points
with the theoretical results obtained in \cite{SS11}. The scalings of 
the ordinates have been chosen for compatibility with the functional form of Eqn.~(60) in \cite{SS11}.
However, it should be noted that 
we have performed simulations for values of $(-\sh \re)$ 
upto about $0.7$, whereas \cite{SS11} have been able to explore larger values of
$(-\sh \re)$.
The plots in Fig.~(\ref{tfnij_th_Pr1}a--c) 
are for $\pr = 1$, but for two sets of values of the Reynolds numbers; 
$\re=\rem \approx 0.16$ (the `bold' lines represent the theory and 
the symbols `$\circ$' represent the simulations),
and $\re=\rem \approx 0.46$ (the `dashed' lines represent the
theory and the symbols `$\times$' represent the simulations).
Figure~(\ref{tfnij_th_Pr5}a--c) are for 
$\re \approx 0.13$ and $\rem \approx 0.64$, corresponding to $\pr \approx 5$ (the
`bold' lines represent the theory and the symbols `$\circ$'
represent the simulations). Figure~(\ref{tfnij_th_Prpt2}a--c) are for $\re \approx 0.13$ and $\rem \approx 0.025$, corresponding to 
${\pr} \approx 0.2$ (the `bold' lines represent the theory and 
the symbols `$\circ$' represent the simulations). Some noteworthy properties are as follows:
\begin{enumerate}
\item[(i)] As may be seen from Fig.~(\ref{tfnij_th_Pr1}), the symbols
`$\circ$' and `$\times$' (also the bold and dashed lines) lie very 
nearly on top of each other. This implies that $\eta_t/(\eta_T\re^2)$, 
$\eta^{\infty}_{12}/(\eta_T\re^2)$ and $\eta^{\infty}_{21}/(\eta_T\re^2)$ are (approximately) functions 
of $\left(-\sh \re\right)$ and $\pr$. Therefore the magnitude of $\chi$ in
Eqn.~(60) of \cite{SS11} should be much smaller than unity. This was 
predicted in \cite{SS11}, and thus our numerical findings are in good agreement
with the theoretical investigations of \cite{SS11}.

\item[(ii)] We see that $\eta_t$ is always positive. For a fixed value of $(-\sh \re)$
the quantity $\eta_t/(\eta_T\re^2)$ increases with $\pr$, and for a fixed value of
$\pr$, it slowly increases with $(-\sh \re)$ (which is consistent with \cite{BRRK08}). 
An excellent agreement between our numerical findings and the theory presented
in \cite{SS11} may be seen from top panels of
Figs.~(\ref{tfnij_th_Pr1}--\ref{tfnij_th_Prpt2}).

\item[(iii)] The quantity $\eta^{\infty}_{12}$ approaches the value
zero in the limit when $(-\sh \re)$ is nearly zero. In the numerical
simulation, it is seen to be
increasing with $(-\sh \re)$ for a fixed
value of $\pr$, and for a fixed value of $(-\sh \re)$ it increases with $\pr$. 
$\eta^{\infty}_{12}$ is expected to behave in a more complicated way. 
Different signs of $\eta^{\infty}_{12}$ are reported in \cite{BRRK08} and \cite{RK06},
whereas both signs have been predicted in calculations of \cite{SS11}. The differences between
the theory and the simulations may be inferred from panels~(b) of
Figs.~(\ref{tfnij_th_Pr1}--\ref{tfnij_th_Prpt2}). 

\item[(iv)] As may be seen from the bottom panels of
Figs.~(\ref{tfnij_th_Pr1}--\ref{tfnij_th_Prpt2}), that, 
$\eta^{\infty}_{21}$ is always positive. This agrees with the
results obtained in earlier works \citep{BRRK08,RS06,RK06}.
Once again, the agreement between our numerical findings and the theoretical
investigations of \cite{SS11}, for this \emph{crucial} component of the diffusivity
tensor is remarkably good\footnote{As discussed in \cite{SS11}, the sign
of $\eta^{\infty}_{21}$ has a direct bearing on the shear--current effect, and
this being positive suggests that the shear--current effect cannot be responsible
for dynamo action, at least in the range of parameters explored.}.
\end{enumerate}

\begin{figure}%[t]
\begin{center}
\includegraphics[scale=0.5]{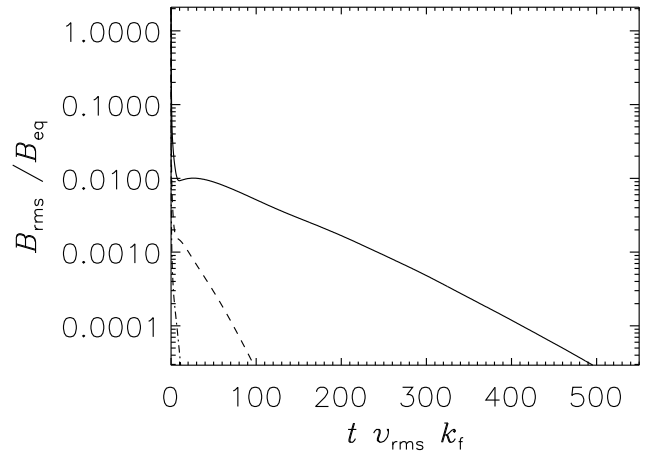}
\caption{Time dependence of the root--mean--squared value 
of the total magnetic field [scaled with respect to $B_{\rm eq}$]
versus the dimensionless parameter $(t\, v_{\rm rms}\, k_f)$. The bold line
is for $\re \approx 0.128$, $\rem \approx 0.643$ (i.e. $\pr=5.0$), 
and $\sh \approx -1.545$; the dashed line is for $\re \approx 0.16$, 
$\rem \approx 0.16$ (i.e. $\pr=1.0$), and 
$\sh \approx -1.237$; and the dashed--dotted line is for $\re \approx 0.127$, 
$\rem \approx 0.025$ (i.e. $\pr=0.25$), and $\sh \approx -1.560$.
$k_f/K = 10.03$ for all three cases.}
\label{lowrelowrm_temp}
\end{center}
\end{figure}

\noindent
Further, we show the time dependence of root--mean--squared value 
of the total magnetic field ($B_{\rm rms}$) in Fig.~(\ref{lowrelowrm_temp}), 
which explicitly demonstrates
the \emph{decay} of $B_{\rm rms}$ for following three sets of 
values of control parameters:
(i) $\re \approx 0.128$, $\rem \approx 0.643$ (corresponding to $\pr \approx 5.0$;
shown by the bold line), 
$\sh \approx -1.545$; (ii) $\re \approx 0.16$, $\rem \approx 0.16$ 
(corresponding to $\pr \approx 1.0$; shown by the dashed line), $\sh \approx -1.237$; 
and (iii) $\re \approx 0.127$, 
$\rem \approx 0.025$ (corresponding to $\pr \approx 0.25$; shown by the
dashed-dotted line), $\sh \approx -1.560$.
Results shown in Fig.~(\ref{lowrelowrm_temp}) are from a direct
numerical simulation with $64^3$ mesh points and $k_f/K = 10.03$.

\begin{center}
 {\bf PART B: $\re > 1$ and $\rem < 1$}
\end{center}

\begin{table*}
\begin{center}
\caption{\label{run_summ_B}
Summary of the simulations for $\re > 1$ and $\rem < 1$}
\begin{tabular}{ccccccccccc}
\hline\hline
\ssc Run &
\ssc $\re$ &
\ssc $\rem$ &
\ssc $k_f/K$ &
\ssc ${-\sh}$ &
\ssc ${\rm Ma}$\footnote{Mach Number} &
\ssc \textrm{Grid} &
\ssc $\eta_t/(\eta_T\re^2)$ &
\ssc $\eta_{12}/(\eta_T\re^2)$ &
\ssc $\eta_{21}/(\eta_T\re^2)$ &
\ssc \textrm{Comments} \\
\hline
A & 5.50 & 0.14 & 10.03 & 0.136 & 0.110 & $64^3$ & 0.000366 & 0.000044 & 0.0000279 & No dynamo \\
B & 4.63 & 0.70 & 10.03 & 0.014 & 0.139 & $64^3$ & 0.006845 & 0.000174 & 0.0000764 & No dynamo \\
C & 4.69 & 0.70 & 10.03 & 0.057 & 0.141 & $64^3$ & 0.007000 & 0.000628 & 0.0003048 & No dynamo \\
D & 4.83 & 0.73 & 10.03 & 0.103 & 0.145 & $64^3$ & 0.007247 & 0.001224 & 0.0005103 & No dynamo \\
E & 5.63 & 0.84 & 10.03 & 0.141 & 0.169 & $64^3$ & 0.006654 & 0.002330 & 0.0006008 & No dynamo \\
F & 41.14 & 0.82 & 3.13 & 0.186 & 0.258 & $64^3$ & 0.000110 & 0.000017 & 0.0000092 & No dynamo \\
G & 48.40 & 0.41 & 3.13 & 0.186 & 0.258 & $64^3$ & 0.000025 & 0.000003 & 0.0000021 & No dynamo \\
\hline
\end{tabular}
\end{center}
\end{table*}

\begin{figure}
\begin{center}
\includegraphics[scale=0.5]{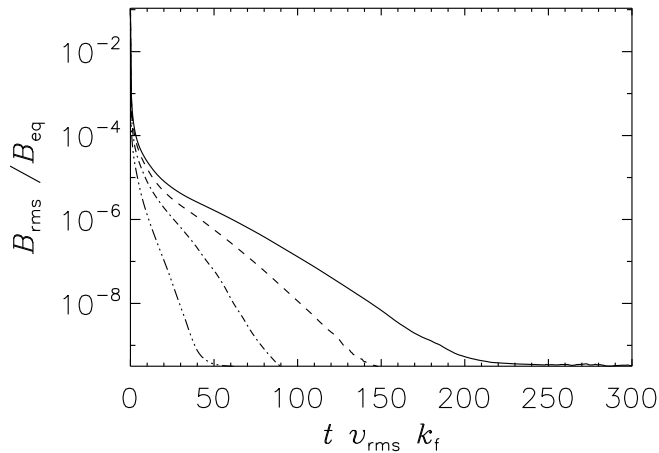}
\caption{Time dependence of the root--mean--squared value 
of the total magnetic field [scaled with respect to $B_{\rm eq}$]
versus the dimensionless parameter $(t\, v_{\rm rms}\, k_f)$. The bold line
is for $\re \approx 24.57$, $\rem \approx 0.614$ (i.e. $\pr=0.025$), 
$k_f/K = 5.09$ and $\sh \approx -0.118$; the dashed line is for $\re \approx 22.40$, 
$\rem \approx 0.448$ (i.e. $\pr=0.02$), $k_f/K = 5.09$ and 
$\sh \approx -0.128$; the dashed--dotted line is for $\re \approx 43.17$, 
$\rem \approx 0.863$ (i.e. $\pr \approx 0.02$), $k_f/K = 3.13$ and 
$\sh \approx -0.177$; and the dashed--dots line is for $\re \approx 36.54$, 
$\rem \approx 0.365$ (i.e. $\pr=0.009$), $k_f/K = 3.13$ and 
$\sh \approx -0.209$}.
\label{fig_highRe}
\end{center}
\end{figure}

We explored this parameter regime for completeness in order to
investigate the dynamo action when $\rem < 1$ whereas $\re > 1$.
Kinematic theory of shear--dynamo problem
was developed in \cite{SS10}, which is valid for low magnetic
Reynolds number but places no restriction on the fluid Reynolds
number. We computed all components of $\alpha_{ij}$ and $\eta_{ij}$
using test--field method and investigated the possibility
of dynamo action. We find that all components of $\alpha_{ij}$ show
fluctuations in time with mean zero and therefore we do not expect
generation of any net helicity in the flow in these
parameter regimes. We summarize all our results 
for $\re > 1$ and $\rem < 1$ in detail in Table~\ref{run_summ_B}.

We find no evidence of dynamo action in this particular parameter regime. This is shown clearly in
Fig.~(\ref{fig_highRe}), in which
we plot the time dependence of root--mean--squared value 
of the total magnetic field ($B_{\rm rms}$) and demonstrate the absence of
dynamo action in this parameter regime. Figure~(\ref{fig_highRe}) shows results from
direct simulation with $64^3$ mesh points for the following four sets 
of parameter values: (i) $\re \approx 24.57$, $\rem \approx 0.614$, 
$k_f/K = 5.09$, $\sh \approx -0.118$ (shown by the bold line);
(ii) $\re \approx 22.40$, 
$\rem \approx 0.448$, $k_f/K = 5.09$, $\sh \approx -0.128$ (shown by the dashed line);
(iii) $\re \approx 43.17$, 
$\rem \approx 0.863$, $k_f/K = 3.13$, $\sh \approx -0.177$ (shown by the 
dashed--dotted line);
and (iv) $\re \approx 36.54$, $\rem \approx 0.365$, $k_f/K = 3.13$, 
$\sh \approx -0.209$ (shown by the dashed--dots line).

\begin{center}
 {\bf PART C: $\re < 1$ and $\rem > 1$}
\end{center}

\begin{figure}
\begin{center}
\includegraphics[scale=0.40]{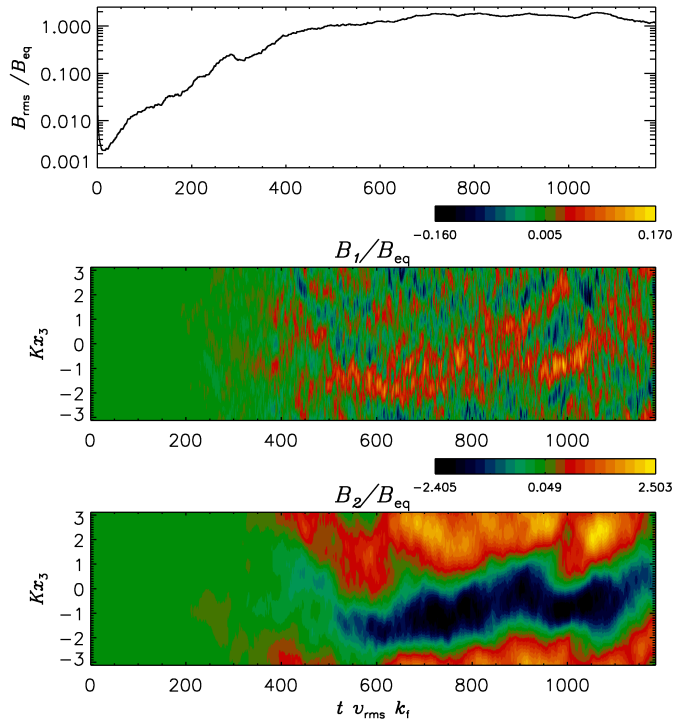}
\caption{Time dependence of the root--mean--squared value 
of the total magnetic field $\bfB^{\rm tot}$ and spacetime diagrams of 
$B_1(x_3, t)$ and $B_2(x_3, t)$ [all scaled with respect to $B_{\rm eq}$] 
from a direct simulation with $\re \approx 0.378$, $\rem \approx 15.135$ 
(i.e. $\pr \approx 40.0$), $k_f/K = 3.13$ and $\sh \approx -1.01$, 
versus the dimensionless parameter $(t\, v_{\rm rms}\, k_f)$. The top 
panel shows the initial exponential growth of the mean magnetic field which 
saturates subsequently with time. The other two panels demonstrate
the episodes of large scale feature in the $x_3-$direction, especially in the
$B_2$ component.\label{DST1}}
\end{center}
\end{figure}

We now report our analysis concerning the growth of mean magnetic
field in a background linear shear flow, with non--helical forcing
at small scale, for the case when $\re < 1$ and $\rem > 1$.
This is a particularly interesting regime for the
following reasons: (i) it is an important fact to note that in the
limit of small $\re$ the non--helical forcing has been shown to give
rise to non--helical velocity field 
(see the discussion below Eqn.~(46) of \cite{SS11});
(ii) For low $\re$ the Navier--Stokes Eqn.~(\ref{NS}) can be linearized
and thus it becomes analytically more tractable problem, as compared
to the case of high $\re$. Such solutions have been rigorously obtained
without the Lorentz forces and have been
presented in \cite{SS11}.
So it appears more reasonable to develop 
a theoretical framework in the limit, $\re < 1$ and $\rem > 1$ before
one aims to have a theory which is valid for both ($\re$, $\rem$) $> 1$.
Such thoughts motivated us to perform numerical experiment in this limit
to look for the dynamo action.
Figures~(\ref{DST1}--\ref{DST3}) display the time dependence of
root--mean--squared value of mean 
magnetic field $\bfB$ and spacetime diagrams of $B_1(x_3, t)$ and
$B_2(x_3, t)$ for three different combinations of $\re$ and $\rem$.
These simulations were performed with $128^3$ mesh points.
We have scaled the magnetic fields in Figs.~(\ref{DST1}--\ref{DST3})
with respect to $B_{\rm eq}$ where 
$B_{\rm eq} = (\mu_0 \langle \rho v_{\rm rms}^2 \rangle)^{1/2}$.
Scalings in these Figures have been chosen for compatibility with
Figs.~(7) and (8) of \cite{BRRK08}. 
Below we list few useful points related to the dynamo action when
$\re < 1$ and $\rem > 1$ based on careful investigation of
Figs.~(\ref{DST1}--\ref{DST3}):

\begin{figure}
\begin{center}
\includegraphics[scale=0.40]{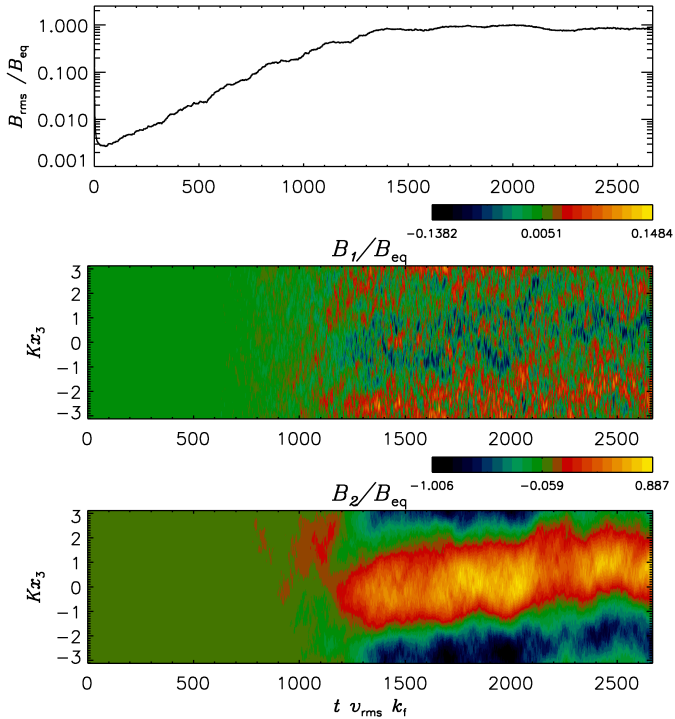}
\caption{Same as Fig.~(\ref{DST1}) but for $\re \approx 0.833$, $\rem \approx 24.976$
(corresponding to $\pr \approx 30.0$), $k_f/K = 3.13$ and $\sh \approx -0.23$.\label{DST2}}
\end{center}
\end{figure}

\begin{figure}
\begin{center}
\includegraphics[scale=0.40]{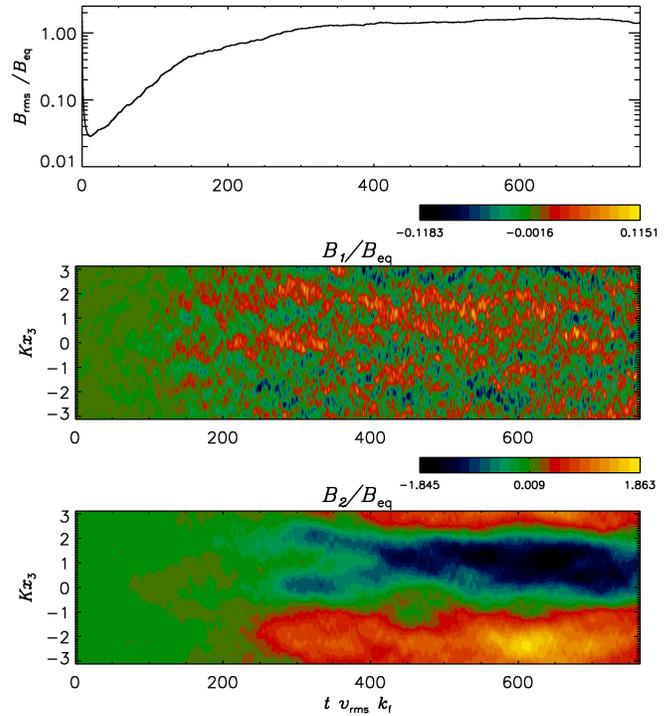}
\caption{Same as Fig.~(\ref{DST1}) but for
$\re \approx 0.641$, $\rem \approx 32.039$ 
(corresponding to $\pr \approx 50.0$), $k_f/K = 5.09$ and
$\sh \approx -0.60$.\label{DST3}}
\end{center}
\end{figure}

\begin{enumerate}
\item [(i)] Top panels of Figs.~(\ref{DST1}--\ref{DST3}) clearly 
show the \emph{growth} of $B_{\rm rms}$ demonstrating the shear dynamo
due to non--helical forcing
($B_{\rm rms}^2 = \left<B^2\right> + \left<b^2\right>$, 
where $B$ and $b$ are the magnitudes of the mean and fluctuating
magnetic fields respectively). Thus the 
$B_{\rm rms}-$field may grow either due to $B$ or $b$, or due to
both $B$ and $b$.
\item [(ii)] Denoting the \emph{magnetic diffusion time scale} as 
$\tau_{\eta} = (\eta k_f^2)^{-1}$ and \emph{eddy turn over time scale} as
$\tau_{\rm edd} = (v_{\rm rms} k_f)^{-1}$, we write 
$\tau_{\eta} = \rem \, \tau_{\rm edd}$. The magnetic fields in these 
simulations survive for times, say $t = 640\, \tau_{edd}$, which for 
$\rem \approx 32$ (corresponding to Fig.~(\ref{DST3})) implies,
$t \approx 20\, \tau_{\eta}$, i.e., twenty times the diffusion time
scale. This is a clear indication of the dynamo action as the
magnetic fields survive much longer than the magnetic diffusion time scale.
\item [(iii)] Spacetime diagrams in Figs.~(\ref{DST1}--\ref{DST3})
reveal that the mean magnetic fields start developing only after times which are
few times the magnetic diffusion time scale ($\tau_{\eta}$).
\item [(iv)] Although the mean magnetic field starts developing at much
later times, $B_{\rm rms}$ starts growing at earlier times. The possibility of
the growth of mean--squared field, with no net mean magnetic field
at these early times, cannot be ruled out.
\end{enumerate}

\begin{table*}%[h!]
\begin{center}
\caption{\label{run_summ_C}
Summary of the simulations for $\re < 1$ and $\rem > 1$}
\begin{tabular}{ccccccccc}
\hline\hline
\ssc Run &
\ssc $\re$ &
\ssc $\rem$ &
\ssc ${-\sh}$ &
\ssc ${\rm Ma}$ &
\ssc \textrm{Grid} &
\ssc $\eta_t/(\eta_T\re^2)$ &
\ssc $\eta_{12}/(\eta_T\re^2)$ &
\ssc $\eta_{21}/(\eta_T\re^2)$ \\
\hline
R1 & 0.393 & 9.837 & 0.391 & 0.10 & $128^3$ & 3.657 & 0.644 & 0.011 \\
R2 & 0.396 & 19.796 & 0.583 & 0.10 & $144^3$ & 4.735 & 1.695 & 0.057 \\
\hline
S1 & 0.592 & 5.920 & 0.325 & 0.15 & $144^3$ & 1.508 & 0.340 & 0.059 \\
S2 & 0.598 & 14.938 & 0.515 & 0.15 & $144^3$ & 2.110 & 0.704 & 0.084 \\
S3 & 0.598 & 29.875 & 0.515 & 0.15 & $144^3$ & 2.400 & 0.786 & 0.140 \\
S4 & 0.603 & 37.692 & 0.638 & 0.15 & $144^3$ & 2.435 & 1.196 & 0.189 \\
\hline
\end{tabular}
\tablecomments{All runs have $k_f/K = 5.10$.}
\end{center}
\end{table*}

{\bf
In Table~\ref{run_summ_C} we present results from test--field simulations
performed in the regime $\re<1$ and $\rem>1$. We have runs for two sets of
values of fluid Reynolds number, $\re$; runs~R1, R2 with $\re \approx 0.4$
and runs~S1--S4 with $\re \approx 0.6$. 
We confirm that all the components of the magnetic diffusivity tensor,
$\eta_t$, $\eta_{12}$ and $\eta_{21}$, increase with increasing magnetic
Reynolds number, $\rem$, which is in agreement with \cite{BRRK08}.
Comparing the numerical values of $\eta$-tensor
in Table~\ref{run_summ_C} (with $\rem>1$) to those in Table~\ref{run_summ_B}
(with $\rem<1$), we see that each component of $\eta_{ij}$ increases
with $\rem$, for range of values considered in this work. For
larger values of $\rem$ we refer the reader to \cite{BRRK08}
where the possibility of the relevant component, $\eta_{21}$,
becoming negative at much larger $\rem\,(> 100)$ was discussed,
although the error bars were quite large, and therefore no conclusion
could be drawn regarding the mean--field dynamo action.
As mentioned earlier, our interest is in the intermediate
$\rem$ values (5--40),
and we show our findings with maximum of $\rem$ being below $38$.
\emph{We note that the component $\eta_{21}$ remains
positive even in this parameter regime}, thus confirming earlier claims
that the shear--current effect cannot be responsible for the observed
large--scale dynamo action. 
}

\begin{figure*}[t!]
\begin{center}
\includegraphics[scale=0.7]{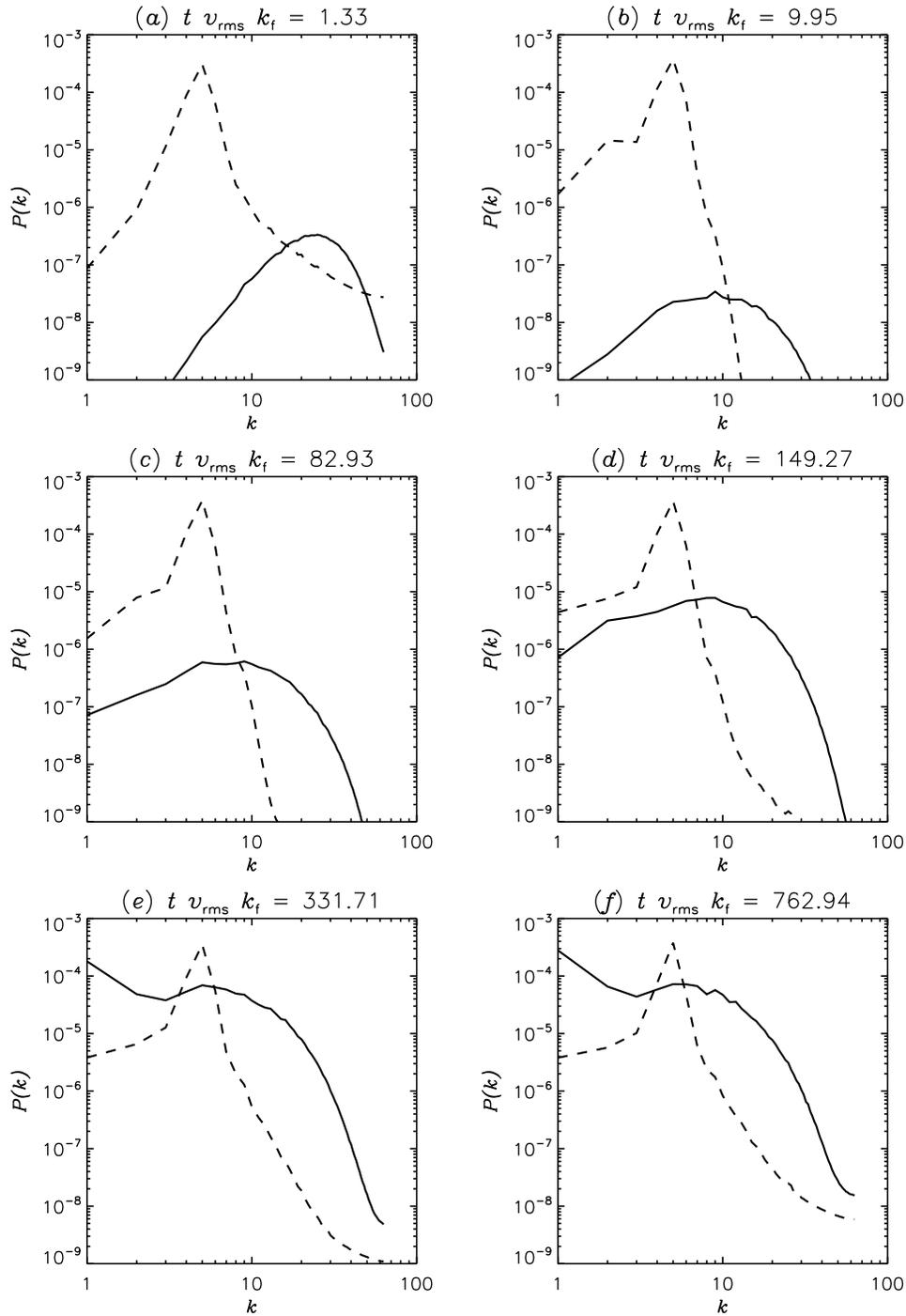}
\caption{Panels ($a$--$f$) show magnetic (bold line) and kinetic
(dashed line) energy spectra from the direct simulation presented
in Fig.~(\ref{DST3}) with $\re \approx 0.641$, $\rem \approx 32.039$,
$k_f/K = 5.09$ and $\sh \approx -0.60$ for different values of
$(t\, v_{\rm rms}\, k_f)$.
\label{ps}}
\end{center}
\end{figure*}

It is instructive to know the magnitude of magnetic power at different
length scales in the simulations and study its evolution in time.
Although the forcing is done at a \emph{single} length scale, a typical
kinetic energy spectrum has a peak at the stirring scale with
significantly less power at other length scales
(e.g., see dashed lines in various panels of Fig.~(\ref{ps})). We display 
in Fig.~(\ref{ps}) the energy spectra obtained in one of the 
three simulations (for different combinations of the control 
parameters, all with $\re < 1$), corresponding to the one shown in
Fig.~(\ref{DST3}). Thus Figs.~(\ref{DST3}) and (\ref{ps}) show results
obtained from one particular simulation with $128^3$ mesh points,
$\re \approx 0.641$, $\rem \approx 32.039$, $k_f/K = 5.09$ and
$\sh \approx -0.60$. A few noteworthy points are discussed below
in detail:

\begin{enumerate}
\item [(i)] Initially the magnetic power is very small as compared to the
kinetic power and it is mainly concentrated at large $k$ (i.e. 
small length scales), as may be seen from panel ($a$) of Fig.~(\ref{ps}).
Also, there is essentially no magnetic power at small $k$
(i.e. large length scales) at the initial stage of the simulation.
\item [(ii)] The strength of the total magnetic field decreases
upto certain time due to dissipation (compare panels ($a$) and
($b$) of Fig.~(\ref{ps})), before it starts building up due to
dynamo action. 
\item [(iii)] From the top panel of Fig.~(\ref{DST3}), we see that
the root--mean--squared value of the total magnetic field starts growing
due to dynamo action ($B_{\rm rms}^2 = \left<B^2\right> + \left<b^2\right>$, 
where $B$ and $b$ are the magnitudes of the mean and fluctuating
magnetic fields respectively). As the $B_{\rm rms}-$field may grow
either due to $B$ or $b$, or due to both $B$ and $b$,
it seems necessary to understand this in more detail.
From Fig.~(\ref{ps}), it may be seen that the magnetic energy
grows at all scales if it starts growing up, till it saturates. 
\item [(iv)] The small scale field grows faster, which
averages out to zero, and hence does not show up in the spacetime
diagrams of Fig.~(\ref{DST3}).
This is generally referred to as the fluctuation dynamo.
The growth rate changes and becomes smaller after the fluctuation
dynamo saturates (which happens at $t\, v_{\rm rms}\, k_f \approx 150$
in Fig.~(\ref{DST3}) and the corresponding power spectrum at that
time is shown in panel ($d$) of Fig.~(\ref{ps})). 
\item [(v)] Although there is non--zero magnetic energy in the
large scales when $t\, v_{\rm rms}\, k_f \approx 150$ (see panel ($d$)
of Fig.~(\ref{ps})), we begin to see some features in the spacetime
diagrams of the mean magnetic field (shown in Fig.~(\ref{DST3}))
only beyond $t\, v_{\rm rms}\, k_f \approx 150$.
Thus, it is possible that $\bfB={\bf 0}$ while $\left<B^2\right>$ be finite.
\item [(vi)] The mean magnetic field starts developing beyond 
$t\, v_{\rm rms}\, k_f \approx 150$ (\emph{which is about five
times the magnetic diffusion time scale}) and saturates at 
$t\, v_{\rm rms}\, k_f \approx 330$ (see Fig.~(\ref{DST3})) 
after which the magnetic energy essentially
stops evolving at all length scales, as may be seen from Fig.~(\ref{ps}).
\item [(vii)] When the magnetic energy saturates at some value,
we see significant magnetic power at the largest scale. 
\end{enumerate}

{\bf
We recall that in the kinematic stage, the magnetic field at all length
scales grow at the same rate, i.e., the magnetic spectrum remains
shape invariant \citep{BS05,SB14}. From panel~($b$) to panel~($d$)
of Fig.~(\ref{ps}), the magnetic spectrum evolves in nearly shape
invariant manner. During this kinematic stage, much of the magnetic power
still lies at small scales, but the power at
the largest scales also grows with time.
This initial growth of magnetic energy occurs at
turbulent (fast) time scale. Towards the end of the kinematic
stage, the growth rates of large and small scale magnetic fields
become different due to the back reaction from Lorentz forces.
The small--scale fields saturate, whereas the large-scale
field continues to grow, thus dominating over small--scale fields at much
later times; shown in panels~($e$) and ($f$) of Fig.~(\ref{ps}).
}

\begin{figure}%[t]
\begin{center}
\includegraphics[scale=1.35]{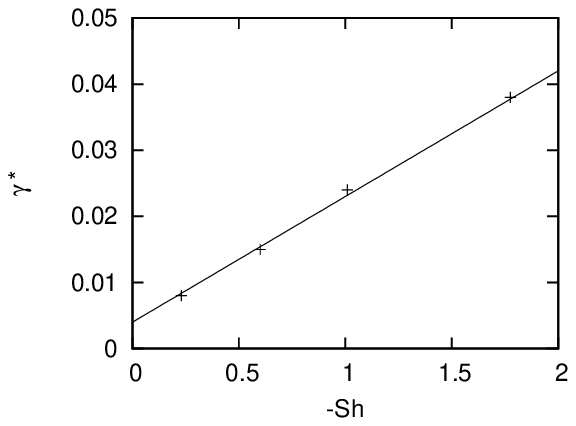}
\caption{Plot of dimensionless initial growth rates, $\gamma^* = \gamma/(v_{\rm rms} k_f)$,
of $B_{\rm rms}$ (corresponding to the cases when $\re < 1$ and $\rem > 1$)
versus $-\sh$. The `$+$' symbols denote results from direct simulations
whereas the bold line shows the slope of the linear trend corresponding to
$\gamma^* \propto -\sh$.\label{grrate}}
\end{center}
\end{figure}

It may be seen from the top panels of Figs.~(\ref{DST1}--\ref{DST3}) that
$B_{\rm rms}$ shows exponential growth. We denote the initial exponential
growth rate of $B_{\rm rms}$ as $\gamma$. It is evident from
Fig.~(\ref{grrate}) that the dimensionless growth rate 
($\gamma^* = \gamma/(v_{\rm rms} k_f)$) appears to scale as
$\gamma^* \propto -\sh$ in the range of
parameters explored in this work. This result is in agreement with 
\citep{You08a,BRRK08,HMS11,RP12}.

\begin{table}%[h]
\begin{center}
\caption{\label{summ_all}
Summary of simulations in different parameter regimes}
\resizebox{\columnwidth}{!}{
\begin{tabular}{cccccccc}
\hline\hline
Run &
$\re$ &
$\rem$ &
$k_f/K$ &
${-\sh}$ &
${\rm Ma}$ &
\textrm{Grid} &
\textrm{Comments} \\
\hline
A1 & 0.47 & 0.47 & 10.03 & 1.27 & 0.0235 & $64^3$ & No dynamo \\
A2 & 0.73 & 0.91 & 10.03 & 0.41 & 0.0727 & $128^3$ & No dynamo \\
A3 & 0.76 & 0.57 & 1.54 & 2.78 & 0.0701 & $128^3$ & No dynamo \\
\hline
B1 & 41.20 & 0.82 & 3.13 & 0.186 & 0.258 & $64^3$ & No dynamo \\
B2 & 4.65 & 0.69 & 10.03 & 0.0285 & 0.139 & $64^3$ & No dynamo \\
B3 & 4.99 & 0.75 & 10.03 & 0.133 & 0.150 & $64^3$ & No dynamo \\
\hline
C1 & 0.59 & 29.47 & 5.09 & 0.12 & 0.03 & $128^3$ & No dynamo \\
C2 & 0.59 & 29.47 & 5.09 & 0.66 & 0.03 & $128^3$ & {\it Dynamo} \\
C3 & 0.85 & 25.50 & 5.09 & 0.226 & 0.129 & $128^3$ & {\it Dynamo} \\
C4 & 0.75 & 33.60 & 10.03 & 0.236 & 0.0674 & $128^3$ & {\it Dynamo} \\
\hline
D1 & 1.04 & 41.66 & 3.13 & 0.367 & 0.13 & $128^3$ & {\it Dynamo} \\
D2 & 1.79 & 89.51 & 5.09 & 0.215 & 0.0911 & $128^3$ & {\it Dynamo} \\
\hline
\end{tabular}
}
\end{center}
\end{table}

In Table~\ref{summ_all} we summarize the details of various simulations
performed in different parameter regimes. We note that larger shear
contributes positively for the mean--field dynamo action; compare the
Runs~C1 and C2, where shear in C2 is 5 times larger compared to C1,
with the rest of the parameters being the same.

\section{Conclusions}

We performed a variety of numerical simulations exploring different regimes of the control parameters
for the shear dynamo problem.
The simulations were done for the following three parameter regimes:
(i) both ($\re$, $\rem$) $< 1$; (ii) $\re > 1$ and $\rem < 1$; and
(iii) $\re < 1$ and $\rem > 1$. These limits, which were never explored in any earlier
works, appeared interesting to us for following reasons: first, to compare analytical findings
of \cite{SS11} with the results of numerical simulations in the
parameter regimes when both ($\re$, $\rem$) $< 1$; and second, to look for the growth of mean magnetic field
in the limit when $\re < 1$.
Exploring the possibility of dynamo action when $\re < 1$ seems
particularly interesting, as, in the limit of small $\re$, non--helical 
forcing has been shown to give rise to non--helical velocity fields
(see the discussion below Eqn.~(46) of \cite{SS11}); whether
this is true even in the limit of high $\re$ has not been proved yet.
Thus performing the simulation in this limit (i.e., $\re<1$) with
non--helical forcing guarantees the fact that \emph{the fluctuating
velocity field is also non--helical}. Also, for low $\re$,
the Navier--Stokes Eqn.~(\ref{NS}) can be linearized and thus it becomes an
analytically more tractable problem, as compared to the case of high $\re$. 
Such solutions have been rigorously obtained without the Lorentz forces, and
have been presented in \cite{SS11}.

In the present paper, we successfully demonstrated that dynamo
action is possible in a background linear shear flow due to non--helical
forcing when the magnetic Reynolds number is above unity whereas the fluid
Reynolds number is below unity, i.e., when $\re < 1$ and $\rem > 1$ (see
Figs.~(\ref{DST1}--\ref{ps})).
Few important conclusions may be given as follows:
\begin{enumerate}
\item We did not find any dynamo action in the limit when both
($\re$, $\rem$) $< 1$ (see Fig.~(\ref{lowrelowrm_temp})). We note
that all the simulations were performed in a fixed cubic domain of
size $2\pi^3$, and the average outer scales of turbulence
in these models were always about ten times smaller than the
domain; see Section~3, part~A. This scale separation of factor
ten might not yet be sufficient, in principle, and the
growth at scales larger than the $x_3$-extent cannot be ruled out.
We computed all the transport coefficients by test--field simulations and
compared with the theoretical work of \cite{SS11} (see
Figs.~(\ref{tfnij_th_Pr1}--\ref{tfnij_th_Prpt2})). A good agreement between the 
theory and the simulations was found for all components of the magnetic diffusivity
tensor, $\eta^{\infty}_{ij}$, except for $\eta^{\infty}_{12}$, which is expected to
behave in a complicated fashion \citep{BRRK08,RK06,SS11}.
\item $\eta^{\infty}_{21}$ was always found to be positive in all the simulations
performed in different parameter regimes. This is in agreement with
earlier conclusions that the shear--current effect cannot be responsible for
dynamo action.
\item There was no evidence of dynamo action in the limit when $\re > 1$ and
$\rem < 1$ (see Fig.~(\ref{fig_highRe})).
\item We demonstrated dynamo action when $\re < 1$ and $\rem > 1$ (see Figs.~(\ref{DST1}--\ref{ps})).
The initial exponential growth rate of $B_{\rm rms}$, $\gamma$, seems to scale \emph{linearly} with
the rate of shear, $|S|$, in the range of parameters explored in this paper
(see Fig.~(\ref{grrate})); a result which is in agreement with \citet{You08a,BRRK08,HMS11,RP12,SS14}.
\end{enumerate}

It's an intriguing question, what drives the dynamo action in
the \emph{non-helical} turbulence. It has been successfully demonstrated,
both, from theoretical and numerical works, that the
\emph{shear-current effect} cannot not be responsible for the
observed shear dynamo. In 1976, Kraichnan discussed the possibility
of zero--mean $\alpha$ fluctuations, which, together with large scale shear,
could possibly give rise to dynamo action in non--helically
forced turbulence \citep{VB97,Sok97,Sil00,Pro07}.
This is known as \emph{the incoherent alpha--shear} mechanism.
{\bf
\cite{HMS11} have predicted the growth of magnetic energy in a shearing
background due to zero--mean $\alpha$ fluctuations, and have obtained
scaling relations in agreement to the results from numerical simulations.
}
In a recent analytical study, \cite{SS14} have shown that the
growth of mean magnetic field is possible due to fluctuating alpha
with non--zero correlation times, in a shearing background.
They derive the dimensionless parameters controlling the nature
of dynamo (or otherwise) action. 
Numerical computation of these
dynamo numbers in simulations of the shear dynamo is being the
focus of a future investigation.

\acknowledgments

We are grateful to S. Sridhar (RRI) for supervising this whole project.
NKS thanks A. Brandenburg (NORDITA) for related discussions over the last
three years and for the hospitality during the visit to NORDITA
in January 2010, where this work began.
We thank K. Subramanian (IUCAA) for suggesting us to
include the power spectrum (given in Fig. 9) and related discussions.
We are grateful to M. Rheinhardt (NORDITA) and the referee for many
useful suggestions. We thankfully acknowledge the cluster facilities
at RRI, IUCAA and NORDITA, where the computations were performed. We
thank Tarun Deep Saini (IISc) for his interest and encouragements.
NJ acknowledges the financial support received by CSIR India.  

\bibliography{ref}
\end{document}